\documentclass[aps,prl,twocolumn,showpacs,superscriptaddress]{revtex4}
\usepackage{graphicx}

\begin{document}

\title{Field-induced coupled superconductivity and spin density wave order in the Heavy Fermion compound CeCoIn$_{5}$}

\author{J.~Spehling}
 \affiliation{Institut f$\ddot{u}$r Festk$\ddot{o}$rperphysik, TU Dresden, D-01069 Dresden, Germany}
\author{R.~H.~Heffner}
 \affiliation{Los Alamos National Laboratory, Los Alamos, New Mexico 87545, USA}
\author{J.~E.~Sonier}
 \affiliation{Department of Physics, Simon Fraser University, Burnaby, Canada}
\author{N.~Curro}
 \affiliation{Departments of Physics, University of California, Davis, USA}
\author{C.~H.~Wang}
 \affiliation{Los Alamos National Laboratory, Los Alamos, New Mexico 87545, USA}
 \affiliation{University of California, Irvine, California 92697, USA}
 \author{B.~Hitti}
 \affiliation{Tri-Meson Facility, University of British Columbia, Vancouver, Canada}
\author{G.~Morris}
 \affiliation{Tri-Meson Facility, University of British Columbia, Vancouver, Canada}
\author{E.~D.~Bauer}
 \affiliation{Los Alamos National Laboratory, Los Alamos, New Mexico 87545, USA}
\author{J.~L.~Sarrao}
 \affiliation{Los Alamos National Laboratory, Los Alamos, New Mexico 87545, USA}
\author{F.~J.~Litterst}
 \affiliation{Institut f$\ddot{u}$r Physik der Kondensierten Materie, TU Braunschweig, D-38106, Germany}
\author{H.-H.~Klauss}
 \email{h.klauss@physik.tu-dresden.de}
 \affiliation{Institut f$\ddot{u}$r Festk$\ddot{o}$rperphysik, TU Dresden, D-01069 Dresden, Germany}
%

\date{\today}

\begin{abstract}

The high field superconducting state in CeCoIn$_5$ has been studied
by transverse field muon spin rotation measurements with an applied
field parallel to the crystallographic c-axis close to the upper
critical field $\mu_{0}H_{c2}=4.97$ T. At magnetic fields
$\mu_{0}H\geq4.8$ T the muon Knight shift is enhanced and the
superconducting transition changes from second order towards first
order as predicted for Pauli-limited superconductors. The field and
temperature dependence of the transverse muon spin relaxation rate
$\sigma$ reveal paramagnetic spin fluctuations in the field regime
from 2 T$\leq\mu_{0}H<$ 4.8 T. In the normal state close to $H_{c2}$
correlated spin fluctuations as described by the self consistent
renormalization theory are observed. The results support the
formation of a mode-coupled superconducting and
antiferromagnetically ordered phase in CeCoIn$_5$ for H directed
parallel to the c-axis.

\end{abstract}


\maketitle


In a conventional type-II superconductor the magnetic upper critical
field $H_{c2}$ is determined by the orbital effect, i.e. the
formation of vortices due to orbital screening currents which
increase the kinetic energy of the Cooper pairs. If the material is
layered or if the effective mass of the quasiparticles becomes large
the orbital effect is reduced and superconductivity can be found up
to higher applied magnetic fields. In this case the so-called Pauli
limit for spin-singlet superconductivity is reached when the
magnetic Zeeman energy, $\mu_B H$, overcomes the binding energy of
the Cooper pairs~\cite{Clogston62}. Close to the Pauli limit,
complex quantum ground states like the
Fulde-Ferrel-Larkin-Ovshinnikov (FFLO) superconducting (SC) phase
\cite{fulde} and other modulated sc phases with mixed
singlet-triplet order (pair density wave, PDW) coupled to a spin
density wave (SDW) magnetic order
\cite{Kenzelmann,Agterberg2,Aperis}, are proposed. CeCoIn$_{5}$, a
layered heavy fermion superconductor with a critical temperature
$T_{c}=2.3$ K, is a model system to test these predictions.

The SC state in CeCoIn$_{5}$ in zero or low magnetic fields is
established to be of spin-singlet $d_{x^{2}-y^{2}}$
symmetry~\cite{Izawa}. In high magnetic fields close to $H_{c2}$,
the SC-to-normal phase transition becomes first order at a
temperature $T_{0}$ ~\cite{Bianchi91,Tayama}. Moreover, a number of
experiments provide evidence for a second order phase transition
inside the SC state near $H_{c2}$ for fields applied along the
a-direction~\cite{Radovan,Bianchi91,Capan,Kakuyanagi}. This
high-field SC (HFSC) state was initially considered as the
realization of the FFLO state~\cite{fulde}. The FFLO state is
characterized by a spatial modulation of the SC order parameter
along the field direction perpendicular to the vortex
lines~\cite{Matsuda}, resulting in a periodic array of planes of
normal paramagnetic electrons. However, NMR experiments in the HFSC
state of CeCoIn$_{5}$  for $H\|$a revealed a site-dependent line
broadening at inequivalent In positions upon entering the
low-temperature SC phase from the homogenous SC phase. This was
interpreted as evidence for antiferromagnetic (AF)
order~\cite{Young}. Subsequent neutron diffraction experiments with
fields applied along (1,-1,0) found indeed an incommensurate SDW at
$\textbf{Q}=(0.44,0.44,0.5)$ in the
HFSC region with a small Ce magnetic moment $\sim0.15\mu_{B}$ 
\cite{Kenzelmann}. SDW order and superconductivity simultaneously
disappear at the same upper critical field, indicating a coupling of
the SC and AF order parameters. In this spatially inhomogeneous
superconducting state the SC condensate carries a finite momentum
corresponding to the modulation wave vector of the magnetic order
$q$. The difference between the order parameter in such a 'Q-phase'
and the FFLO order parameter is that in the 'Q-phase' $q$ is $H$
independent and of short wave length, and also that mixed singlet
and triplet SC order parameters can occur.

For the field direction  $H\|$c the existence of such phases in
CeCoIn$_5$ is still under debate and remains a matter of great
interest~\cite{Bianchi91,Kumagai,bianchi08}. In a local probe
experiment like NMR or $\mu$SR, a long-wave-length modulation of the
SC condensate density is expected to cause an additional resonance
line due to the introduced paramagnetic planes~\cite{Ichioka};
therefore such experiments are ideally suited to identify the FFLO
phase. NMR measurements by Kumagai et al. revealed such a double
peak structure only close to $T_{c}$~\cite{Kumagai}. However, this
structure may be caused by RF heating, as demonstrated by Mitrovic
et al.~\cite{Mitrovic2}.

\begin{figure}
\begin{center}
\includegraphics[width=1\columnwidth]{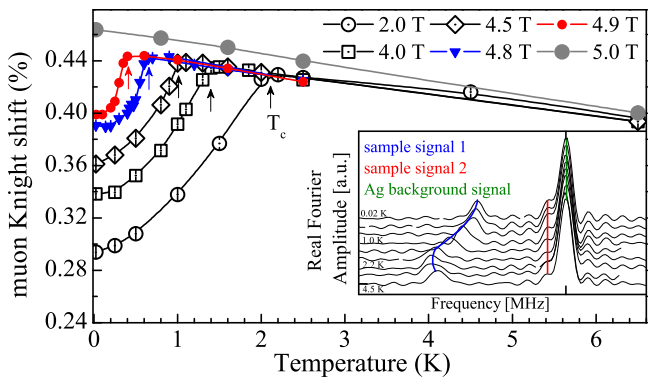}
\end{center}
\caption[1]{(color online). Muon Knight shift as a function of
temperature at various magnetic fields. Arrows indicate the SC
transition temperatures $T_{c}$. Lines are guides to the eye. Inset:
Fourier transformed muon spin precession spectra at different
temperatures at an applied field of 2 T.} \label{fig:fig1}
\end{figure}

In this Letter, we present a detailed transverse field (TF) muon
spin rotation ($\mu$SR) study on single crystalline CeCoIn$_{5}$. We
focus on the low-temperature (0.02 to 7 K) and high-field (2 to 5 T)
part of the $H-T$-phase diagram for $H\|$c. Our experiments reveal a
field-induced enhanced local spin susceptibility close to $H_{c2}$,
which becomes temperature-independent at low temperatures,
consistent with the appearance of a triplet component in the SC
order parameter. Our results further show a strong nonlinear
enhancement of the static line width and the dynamic
$({T_{1}T})^{-1}$ relaxation rate with increasing field below
$H_{c2}$ consistent with a coupled SC and AF order. Our data do not
show the appearance of an additional peak at the paramagnetic
position in the local probe frequency spectrum, as predicted for a
FFLO state with a long wave length modulation $q$.


Single crystals of CeCoIn$_{5}$ were grown in In
flux~\cite{Petrovic}.  The measurements were performed at the M15
beam line, TRIUMF, Vancouver, Canada, using a top loading dilution
refrigerator.  External magnetic fields up to 5 T were applied
parallel to the crystallographic c-axis transverse to the initial
$\mu$-spin polarization. All experiments were performed under field
cooled conditions. The muon Knight shift
$K=\frac{\omega_{\mu}-\omega_{Ag}}{\omega_{Ag}}$ can be expressed as
$K=(A_{c}+A_{dip})\cdot\chi_{4f}+K_{dem}+K_{0}+K_{dia}$. Here,
$A_{c}+A_{dip}$ describes the contact and dipolar coupling of the
muon to the $4f$ susceptibility $\chi_{4f}$, $K_{dem}$ is the
correction for demagnetization and Lorentz fields, $K_{0}$ is the
isotropic temperature-independent contribution from non-$4f$
conduction electrons and $K_{dia}$ is due to flux expulsion in the
SC state. In type-II superconductors the last term can lead to an
internal field distribution $p(B_{local})$, with
$\omega_{\mu}=\gamma_{\mu}B_{local}$ and $\gamma_{\mu}/2\pi =$
135.5342 MHz/T, which can be measured via a contribution to the
static line width $\sigma_{s}$.

\begin{figure}
\begin{center}
\includegraphics[width=1\columnwidth]{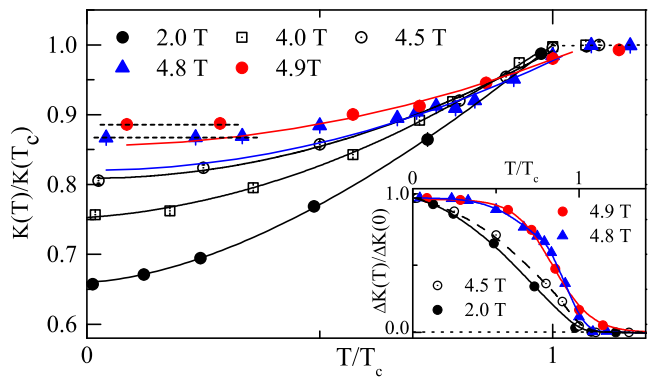}
\end{center}
\caption[1]{(color online). Normalized muon Knight shift
$K(T)/K(T_{c})$ in the SC state as a function of the reduced
temperature. Solid lines denote a quadratic $T$-dependence. Inset:
Normalized muon Knight shift $\Delta K(T)/\Delta K(T\rightarrow0$)
as a function of $T/T_{c}$.}
\end{figure}


The Fourier transformed muon spin precession spectra (inset, fig.~1)
show one distinct symmetric line (sample signal 1) and one
non-symmetric line which is due to the overlap of  sample signal 2
and a temperature independent Ag background signal. Two sample
signals, distinct by their different precession frequencies,
indicate two magnetically inequivalent muon sites. Calculations of
the dipolar hyperfine coupling tensor $A_{dip}$ for localized
Ce-$4f$ moments interacting with the muon spin identify signal 2
with the muon stopping site
($\frac{1}{2}$,$\frac{1}{2}$,$\frac{1}{2}$) and signal 1 with the
muon stopping sites ($\frac{1}{2}$,0,0) and (0,$\frac{1}{2}$,0).
The hyperfine coupling at site 2 is an order of magnitude smaller
than at site 1. Hence, it strongly overlaps with the Ag signal and,
therefore, our discussion will focus solely on signal 1 due to its
clear separation. Moreover, sample signal 2 shows nearly no
frequency shift with respect to the Ag background signal at all
temperatures studied. Since any demagnetization or Lorentz fields as
well as the diamagnetic orbital contribution in the SC state to
$K(T)$ would be the same at all muon sites we conclude that those
contributions are negligible and temperature independent. Therefore,
$K(T)$ directly samples the local spin susceptibility
$\chi_{4f}(T)$.

In the SC phase at all examined field strengths signal 1  is
homogeneous and does not show a splitting with an additional line at
the paramagnetic position as expected for a FFLO
phase~\cite{Ichioka}. The main panel of fig.~1 shows the muon Knight
shift $K(T)$ as a function of temperature for various applied
fields.  As the temperature is lowered towards $T_{c}$, $K(T)$
slightly increases due to a Curie behavior of the spin polarization.
Below $T_{c}$, $K(T)$ reflects the suppression of $\chi_{4f}(T)$ due
to spin singlet pairing. In the case of s-wave spin singlet pairing,
$\chi_{4f}(T)$ is expected to vanish for $T \rightarrow 0$
K~\cite{Yosida}. However, spin-orbit (SO) scattering by impurities
and unconventional pairing states, e.g. d-wave pairing with nodes in
the gap function or spin-triplet pairing can cause $\chi_{4f}(0)$ to
be finite and to increase toward its normal state value. Increasing
the external field suppresses $T_{c}$ towards smaller temperatures.
At an applied of 5 T, superconductivity is completely suppressed. An
upper critical field of $\mu_{0}H_{c2}=4.97$ T is deduced from a fit
to the $H(T_{c})$ data (not shown). Note, the small 4\% increase of
the Knight shift in 5 T external field indicates a slightly enhanced
spin polarization in the paramagnetic state.

The inset of fig.~2 depicts the Knight shift change $\Delta K(T)$
normalized to $\Delta K(T\rightarrow0)$ as function of the reduced
temperature. Whereas a relatively small change occurs from 2 T to
4.5 T, we observe a substantial change between 4.5 T to 4.8 T
indicating a field-induced change from a second order towards a
first order phase transition between the normal and the SC state.
Note that SC fluctuations~\cite{Adachi} may modify the first order
transition into a nearly continuous crossover. The change from
second towards first order appears at a corresponding temperature of
$T_{0}\approx0.65$ K $\approx0.3T_{c}(H=0)$. This agrees well with
reported values~\cite{Bianchi91,Tayama} and the predictions by
Gruenberg and Gunther for Pauli-limited
type-II-superconductors~\cite{Gruenberg}.

The $T$-dependence of the local magnetization in the SC state is
expected to be quadratic for high magnetic fields when
$\mu_{B}B>k_{B}T$ ~\cite{Vorontsov2,Yang}, with a residual value
$\chi_{4f}$~\cite{footnote2}. The main panel of fig.~2 displays
$K(T)$ normalized to $K(T_{c})$ in the SC state. For $\mu_{0}H<4.8$
T the data agree well with the predicted quadratic $T$-dependence.
For $\mu_{0}H\geq4.8$ T we observe a discrepancy. This discrepancy
close to $H_{c2}$ is particularly pronounced at very low
temperatures where the local spin susceptibility is enhanced and
nearly temperature independent (dashed lines) in comparison to the
$\mu_{0}H<4.8$ T data.  This result is consistent with Knight shift
calculations from Aperis et al. considering a field-induced
coexistence of $d$-wave spin-singlet superconductivity with
staggered $\pi$-triplet
superconductivity~\cite{Aperis,Aperis2,Kyung,Psaltakis}. Moreover,
our Knight shift data mirror NMR results showing a comparable
behavior of the spin susceptibility for fields along
a-axis~\cite{Mitrovic1} indicating a common high field ground state
in CeCoIn$_5$ regardless of the direction of the external field.


\begin{figure}
\begin{center}
\includegraphics[width=1\columnwidth]{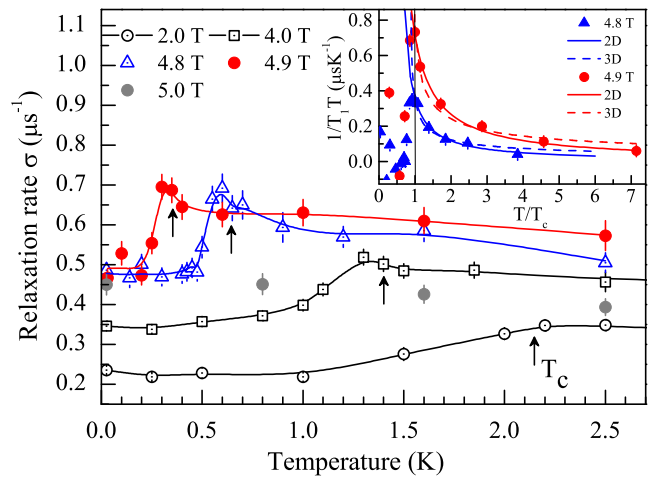}
\end{center}
\caption[1]{(color online). Muon relaxation rate $\sigma$ as a
function of temperature at various magnetic fields. Arrows indicate
$T_{c}$ derived from Knight shift data. Lines are guides to the eye.
Inset: $\frac{1}{T_{1}T}$ vs. $T/T_{c}$ at 4.8 T and 4.9 T. Solid
lines denote fit results according to Moriya's SCR theory. See text
for details.}
\end{figure}

Fig.~3 shows the temperature dependence of the muon spin relaxation
rate $\sigma$ at various applied magnetic fields. For all field
strengths ($\mu_{0}H\geq2 T $) we find a \textsl{reduction} of
$\sigma$ in the SC state with decreasing temperature. This is
contrary to the expected increase of the static line width
$\sigma_{s}$ in the vortex state due to the formation of the flux
line lattice (FLL) and the decrease of the in-plane magnetic
penetration depth $\lambda_{ab}$ in the SC phase, i.e.
$\sigma_{s}\propto1/\lambda^{2}_{ab}$. Note that our results differ
considerably from low-field (0.3 T) TF-$\mu$SR data, which show a
slight \textit{increase} of $\sigma$ in the SC
state~\cite{Higemoto}.

Since there is no common mechanism which leads to a reduction of the
static line width below a conventional SC phase transition we now
consider that the transverse field muon relaxation rate $\sigma$ is
the sum of a static line width $\sigma_{s}$ and a dynamic $1/T_{1}$
contribution. To extract $1/T_{1}$ we assume a
temperature-independent $\sigma_{s}$~\footnote{$\sigma (20 mK)$ is
taken as the static line width $\sigma_{s}$ since we can expect
$\frac{1}{T_{1}}=0$ at 0 K. To deduce $\sigma_{s}$ in the normal
state for fields $\mu_{0}H>4.8$ T we take the linear interpolated
value between $\sigma_{s}$ (2 T) and $\sigma_{s}$ (5 T) since the
unusual enhancement of the static line width around 4.85 T shown in
fig.~4 is a unique feature of the combined superconducting/magnetic
state. Therefore one can assume this enhancement is not present in
the normal state at these magnetic field strengths.}. This is
reasonable since at 5 T external field ($B>B_{c2}$) there is nearly
no change in the relaxation rate with temperature in the $T$-regime
studied. A sizeable dynamic contribution is deduced only for those
field values at which a SC transition is found. Therefore, at 5 T
external field $\sigma$ is decreased in comparison with the data at
lower field values in the paramagnetic regime.

The dynamic muon relaxation rate $\frac{1}{T_{1}}$ measures the
magnetic fluctuation spectrum of the electronic spin system at
frequency $\omega_{\mu}$, i.e., $\frac{1}{T_{1}T}$ is proportional
to the \textbf{q}-integrated dynamic magnetic susceptibility
$\chi''(\omega_{\mu})$.  We plot the extracted dynamic muon
relaxation  depicted as $\frac{1}{T_{1}T}$ in the inset of fig.~3
for applied fields of 4.8 T and 4.9 T and for $T>T_{c}$. For a metal
one expects a Korringa behavior, i.e. $\frac{1}{T_{1}T}$
 to be constant. Instead, we observe that the $T$-dependence of
$\frac{1}{T_{1}T}$ follows a power-law,
\begin{eqnarray}
\frac{1}{T_{1}T}\propto(T-T_{N})^{-\beta}, \qquad T > T_{c},
\end{eqnarray}
diverging at finite temperatures below $T_{c}$. This indicates a
critical slowing down of magnetic spin fluctuations close to a
second order magnetic phase transition below $T_{c}$. The data can
be described equally well with a critical dynamic exponent $\beta =
1$ for 2-D or $\beta = 0.5$ for 3-D AF correlations according to
Moriya's self consistent renormalization (SCR) theory~\cite{Moriya}.
These magnetic correlations induce the dynamic muon spin relaxation
dominating the muon relaxation rate $\sigma$ for $T>T_{c}$. Note, at
all measured field strengths below 5 T, $\frac{1}{T_{1}T}$ can be
described by a power-law indicating the presence of critical spin
fluctuations. Whereas our $\mu$SR data at 2 T and 4 T diverge at
$T=0$ K indicating an AF quantum phase transition~\cite{curro2}, the
high field data show a divergence at finite temperatures below
$T_{c}$ ($T_{N}/T_{c}\approx 0.5 $ assuming 2-D and
$T_{N}/T_{c}\approx 0.9$ assuming 3-D correlations) as if there were
an accompanying magnetic phase transition. In the SC state due to
spin singlet formation a reduction of $\frac{1}{T_{1}T}$ is
observed. Our relaxation rate data agree qualitatively well with the
calculated spin-lattice relaxation rate $\frac{1}{T_{1}T}$ in
Ref.~\cite{Aperis}. This further supports the suggested scenario of
SDW order triggered by the emergence of a $\pi$-triplet pairing
component in the superfluid~\cite{Aperis,Aperis2,Agterberg2}. In
contrast to Ref.~\cite{Aperis} in the normal state we observe an
increase of $1/T_{1}T$ as $T_C$ is approached from higher
temperatures which is attributed to the vicinity to a magnetic phase
transition.


\begin{figure}
\begin{center}
\includegraphics[width=1\columnwidth]{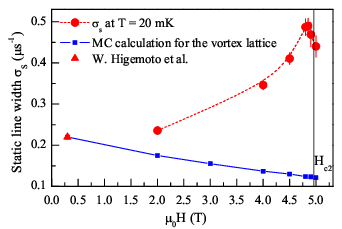}
\end{center}
\caption[1]{(color online). Field dependence of the static line
width $\sigma_s$ at $T = 20$ mK (red circles).  The blue squares
depict calculated values for an Abrikosov vortex lattice with
$\lambda=5500${\AA} and $\xi=46.84${\AA}. Lines are guides to the
eye.}
\end{figure}

The field dependence of the static muon relaxation rate $\sigma_{s}$
at $T=20$ mK is depicted in fig.~4. It shows a non-linear increase
of the internal field distribution in the SC state. This is in
conflict with the Ginzburg-Landau paradigm, which predicts a
monotonically decrease of $\sigma_{s}$ for $H>H_{c1}$, with
increasing field as verified by Monte Carlo
calculations~\footnote{MC calculation of the internal field
distribution due to the vortex lattice formation assuming a Cooper
pair coherence length $\xi = 46.84${\AA} and a London penetration
depth $\lambda = 5500${\AA}~\cite{Higemoto}.}. The low-field data
point~\cite{Higemoto} fits well into this expected field dependence
as denoted by the red triangle.  The strongly enhanced static line
width around 4.85 T is associated with a field-induced magnetic
order
 inside the SC phase. Our findings are
similar to recent neutron diffraction data for $H\|c$, which report
an unusual enhancement of the vortex lattice form factor consistent
with localized spins inside the vortex cores~\cite{bianchi08}. The
decrease of the static line width at 5 T external field implies that
the additional source of magnetism vanishes simultaneously at the
same critical field $H_{c2}$. This confirms the intimate coupling
between magnetism and
superconductivity~\cite{Kenzelmann,Agterberg2,Aperis} in
CeCoIn$_{5}$.


In conclusion, we performed transverse field $\mu$SR measurements at
high magnetic fields up to 5 T with $H\|$[001] to examine the
microscopic nature of the SC state in CeCoIn$_{5}$. The local spin
susceptibility $\chi_{4f}(T)\propto K(T)$ is reduced in the SC state
consistent with spin-singlet superconductivity. Close to $H_{c2}$ we
find an unusual enhanced and temperature independent spin
susceptibility at low temperatures consistent with theoretical
calculations \cite{Aperis} of the appearance of a staggered
spin-triplet SC component coexisting with the $d-$wave singlet
superconductivity. We further observed a strong non-linear increase
of the static line width as well as the dynamic $\frac{1}{T_{1}T}$
relaxation rate with increasing field strength in the SC state.
These results support a field-induced coupled SC and AF phase
transition as described by a pair density wave (PDW) theory
discussed e.g. by Agterberg et al. and Aperis et
al.~\cite{Agterberg2,Aperis}.


This work was performed at the Tri-University Meson Facility
(TRIUMF), Vancouver, Canada. We acknowledge with thanks the help of
D. Arsenau, S. Kreitzman and the TRIUMF accelerator crew. Work at TU
Dresden is supported by the DFG under Grant No. KL1086/8-1. Work at
UC Irvine is supported by U.S. DOE Grant No. DE-FG02-03ER46036. Work
at Los Alamos was performed under the auspices of the U. S.
DOE/Office of Science and supported by the Los Alamos LDRD program.

\end{document}